\pdfoutput=1
\documentclass[
    aps,pre,twocolumn,
    reprint,
    superscriptaddress,
    nofootinbib,
    floatfix,
    amssymb,
    longbibliography
]{revtex4-2}
\usepackage[english]{babel}

\usepackage{amsmath}
\usepackage{amsfonts}
\usepackage{graphicx}
\usepackage{bm}

\usepackage[utf8]{inputenc}

\usepackage{nicefrac}

\usepackage[normalem]{ulem}

\usepackage{xcolor}

\definecolor{linkColor}{rgb}{0,0.3,0.7}
\usepackage{hyperref}
\hypersetup{colorlinks=true,
            allcolors=linkColor,
            pdfborder={0 0 0},
            pdfencoding = auto
            }

\begin{document}

\title{Feedback-induced mechanical bistability and perpendicular-domain formation in active elastic tissues}

\author{Divyoj Singh}
\thanks{D.S.\ and H.W.\ contributed equally to this work.}
\email{divyoj@ucsb.edu}
\affiliation{Department of Physics, University of California Santa Barbara, Santa Barbara, California 93106, USA}

\author{Henrik Weyer}
\email{henrik.weyer@kitp.ucsb.edu}
\affiliation{Kavli Institute for Theoretical Physics, University of California Santa Barbara, Santa Barbara, California 93106, USA}

\author{Harry Kroft}
\email{hkroft@ucsb.edu}
\affiliation{Department of Physics, University of California Santa Barbara, Santa Barbara, California 93106, USA}

\author{M. Cristina Marchetti}
\email{cmarchetti@ucsb.edu }
\affiliation{Department of Physics, University of California Santa Barbara, Santa Barbara, California 93106, USA}

\date{November 14, 2026}
	
\begin{abstract}

Contractile fibers in muscle tissues are highly aligned over the scale of many cells. Muscle layers surrounding organs with an inner lumen are often arranged in orthogonal directions. Here, we study a minimal mechanism of how structural order and its patterning into orthogonally aligned domains can emerge from mechanical feedback alone. The central idea is that contractile fibers align in the direction of the local \emph{active and elastic} stresses, a feedback motivated by the maturation of muscle fibers under tension. We thus model muscle tissue as an active nematic solid with stress feedback: an elastic medium with embedded contractile fibers that are reinforced along the directions of the local tissue stresses. We find that the feedback from the contractile fiber stress is self-amplifying and drives spontaneous fiber alignment. Feedback from the elastic stress couples the alignment to the mechanical boundary conditions and tissue geometry. Depending on the mechanical boundary conditions, the elastic-stress feedback stabilizes two orthogonally aligned tissue states, resulting in a structurally and mechanically bistable tissue that shows a hysteretic stress--strain response. In spatially extended tissues, this bistability gives rise to orthogonally aligned domains separated by domain walls. Our results show that stress feedback can pattern structural order by coupling spontaneous alignment to the tissue geometry. Our mechanism thus provides a purely mechanical route to orthogonal fiber patterning in muscle tissues.

\end{abstract}
\maketitle

\section{Introduction}

During the development of muscle tissue, the contractile machinery of cells self-organizes into remarkable structural order. Muscle fibers almost perfectly align along a common axis over hundreds or thousands of cells, allowing for effective directional force generation. This type of order is found \textit{in vivo} in cardiac, skeletal, and smooth muscle \cite{Hill.Olson2012}, in entire organisms, such as \textit{Hydra} \cite{aufschnaiterApicalBasalEpitheliomuscular2017}, and \textit{in vitro} in engineered skeletal and cardiac muscle tissues  ~\cite{parker2008myofibrillar,fenix2018muscle,mao2022tension}, as well as in cardiac organoids~\cite{Merk.etalinpreparation}.

It has been suggested that this structural organization emerges through mechanical feedback \cite{weitkunat2014tension,yuan2017myofibrils,mao2022tension,baillesAnisotropicStretchBiases2025a, hannezoMechanochemicalFeedbackLoops2019,kolley-kochelPhysicalBasisSarcomere2025}.
Aligned fibers assemble \textit{de novo} from initially unaligned cytoskeletal material over hours to days ~\cite{weitkunat2014tension, fenix2018muscle,yuan2017myofibrils,mao2022tension},
and their assembly, maturation, and orientation are mechanosensitive. Myofibrils, the elongated contractile actomyosin bundles that make muscle contraction possible,  assemble only under tension in developing Drosophila flight muscle~\cite{weitkunat2014tension} and along the directions of maximal principal stress in cardiac muscle cells on micropatterned substrates~\cite{yuan2017myofibrils}. The formation of sarcomeres, the basic contractile unit of mature cardiac and skeletal muscle fibers, is argued to be mediated by the forces transmitted along the fiber~\cite{kolley-kochelPhysicalBasisSarcomere2025}. On the molecular level, such tension feedback may be implemented through force-sensitive molecules, the catch-bond behavior of myosin-actin binding, or the load-dependent reinforcement of focal adhesions and cell--cell junctions \cite{guoMechanicsActomyosinBonds2006, rivelineFocalContactsMechanosensors2001, kolley-kochelPhysicalBasisSarcomere2025, schwarz2013physics}.

In many organs, the structural order of muscle tissue is precisely patterned to allow force generation not only along a single axis but in several directions. Examples are the myocardium of the heart \cite{streeter1969fiber}, the smooth muscle cells of the gut \cite{huyckeGeneticMechanicalRegulation2019} and of the arterial wall \cite{spronck2016method}, and the epithelial fibers of \emph{Hydra}~\cite{aufschnaiterApicalBasalEpitheliomuscular2017,maroudas-sacksTopologicalDefectsNematic2021}. 
In particular, muscle layers that surround organs with an inner lumen, are frequently arranged in orthogonally aligned structures.
For instance,  the intestine of chicken consists of an inner circumferential and an outer longitudinal smooth-muscle layer~\cite{huyckeGeneticMechanicalRegulation2019}. Similarly, the \emph{Drosophila} midgut is enclosed by an inner layer of circular and an outer layer of longitudinal visceral muscles \cite{martinDistinctSetFounders2001}. In \emph{Hydra},
whose body wall encloses a central gastric cavity, the ectodermal actomyosin fibers run along the body axis, while the endodermal ones run
circumferentially~\cite{aufschnaiterApicalBasalEpitheliomuscular2017}. Cardiac organoids with a central lumen similarly develop circumferentially and radially aligned regions~\cite{Merk.etalinpreparation}. It remains unclear whether a generic mechanical-feedback mechanism may generate this type of perpendicular alignment. 

Different theoretical models have been introduced to study the interplay between the alignment of contractile fibers and mechanical feedback. At the subcellular scale, models of sarcomere and myofibril self-assembly explore how contractile units nucleate and order within a single fiber~\cite{kolleyMechanismsSarcomereAssembly2024, kolley-kochelPhysicalBasisSarcomere2025}. At the cellular scale, physical models describe how adherent cells and their stress fibers orient in response to the mechanics of their environment~\cite{schwarz2013physics} and how cell shape and feedback between actin fibers and focal adhesions orient precursory myofibrils~\cite{grosbergSelfOrganizationMuscleCell2011}.
At the tissue scale, extensive previous work has focused on modeling tissue as active nematic fluids where cells flow and rearrange to release stress~\cite{Prost2015}.
Here, in contrast, we seek to describe muscle tissue that structurally orders but does not change shape significantly.
Moreover, cell proliferation and intercalation are negligible in these tissues after an initial growth phase \cite{white2010dynamics, bergmann2015dynamics}.
We therefore describe the large-scale tissue response as elastic and model tissues as active solids.

In the setting of active solids, numerical models of the left ventricle have been used to simulate different fiber-alignment mechanisms based on local tissue strains and stresses~\cite{Arts.etal1994,Kroon.etal2009,Pluijmert.etal2012,Washio.etal2020,Mehri.etal2024}.
Moreover, models of adherent cells that sense the stiffness and strain of their elastic substrate have shown that elastic interactions transmitted through the substrate can drive cell alignment~\cite{bischofs2003cell,bischofs2005effect,Zemel.etal2010,De.etal2007,Zemel.etal2010}.
Coupling to a strain-stiffening substrate, such mechanosensitivity of cells can create bistability in cellular tractions~\cite{pi2026bistability}.
Moreover, the nematic alignment of stress fibers has been modeled as the ordering of force dipoles in an elastic background~\cite{friedrich2011nematic}. At the tissue scale, models where fibers reorient in the direction of principal strain reproduce the alignment of endothelial actin in pressurized vessels~\cite{dessallesInterplayActinNematodynamics2025}.
The folding of engineered cell sheets with prescribed cellular alignment is captured by modeling the tissue as an active contractile elastic sheet~\cite{guillamat2026guidance}.
Conceptually, recent work in active solids, coupling orientational order to elasticity, has studied the onset of (quasi-)long-range order and its instability~\cite{maitraOrientedActiveSolids2019,yin2023contractility,
yangOrientationalOrderingActive2025}, the self-propulsion of topological defects~\cite{braunsActiveSolidsTopological2026}, and soft elasticity and active yielding in feedback-controlled epithelia~\cite{yuFeedbackcontrolledEpithelialMechanics2026}.

These models differ in the mechanical quantity to which the fiber alignment is coupled. In models of adherent cells, cells sense the strain in their elastic environment \cite{bischofs2003cell, bischofs2005effect}. Since this environment is deformed by both external loads and the contraction of the cells themselves, this microscopic strain coupling effectively encompasses, when coarse-graining to the tissue scale, coupling to both the elastic and active stresses \cite{friedrich2011nematic}. Other models reorient the fibers based on the local tissue strain, as for endothelial tubes \cite{dessallesInterplayActinNematodynamics2025} and classical numerical models of the fiber architecture of the left ventricle \cite{Arts.etal1994,Kroon.etal2009,Pluijmert.etal2012,Washio.etal2020}.
Given the diverse mechanisms of mechanosensitivity discussed above, and local tension created by both active and elastic stresses, we consider a generic coupling of the fiber orientation to active and elastic stresses, where the elastic stress is related to the tissue strain.
When coupling equally to active and elastic stress, this coupling becomes a coupling to the total stress \cite{Dye.etal2021,Ioratim-Uba.etal2023,ioratim2025exotic,Mehri.etal2024}. The generic effects of stress-driven self-organization in active elastic tissues remain unknown.

In this paper, we examine a minimal continuum model of active elastic tissues with stress feedback, extending our brief report in Ref.~\cite{Letter2026}.
We describe the tissue as a linearly elastic material with embedded contractile fibers in which fiber remodeling is driven by mechanical feedback from \emph{both} elastic and active-fiber stress. We show that the feedback from fiber contractility is self-amplifying and leads to spontaneous fiber alignment throughout the tissue without any external cue. Feedback from the elastic stress breaks the rotational degeneracy of the ordered state as the mechanical boundary conditions select two special directions, parallel and perpendicular to the applied load.
The stability of these two possible orientations depends on geometry and boundary conditions.
Fiber alignment parallel to the load is always stable, whereas fiber alignment normal to the load is only stable if the boundary conditions allow the tissue to shear.
In this case, the tissue is structurally and mechanically bistable, providing a minimal mechanism for orthogonal fiber order.
In the spatially extended setting, this bistability can pattern the tissue into orthogonally aligned domains separated by domain walls.
While the domain wall generically acquires a finite velocity in quasi one-dimensional tissue, it may be pinned in biologically relevant tissue geometries.

The manuscript is structured as follows. In Sec.~\ref{sec:model}, we introduce the model of an active elastic tissue with stress feedback. In Sec.~\ref{sec:steady-states}, we analyze the homogeneous steady states of this model tissue, highlighting the roles of feedback from the active fiber stress  (Sec.~\ref{sec:chie0}) and from the elastic stress under different mechanical boundary conditions (Sec.~\ref{sec:chie}). In Sec.~\ref{sec:lsa}, we carry out a linear stability analysis of these states. In Sec.~\ref{sec:potential} we show that the uniform dynamics can be mapped onto a gradient flow in an effective potential, summarizing the findings on structural and mechanical bistability. In Sec.~\ref{sec:quasi-1D-domain-wall} we turn to spatially extended tissues and the formation of domain walls in a quasi-one-dimensional tissue patch. 

A concise account of the self-alignment transition 
and the resulting mechanical hysteresis
is given in the accompanying Letter~\cite{Letter2026}. Here, we present the analytical treatment and broader behavior of active elastic tissues under different mechanical boundary conditions. 
 
\section{Active-solid model with stress feedback}
\label{sec:model}
In this paper, we examine in detail the model introduced in Ref.\cite{Letter2026} of a biological tissue as a solid-like active nematic.
The model describes the tissue as composed of an elastic background (consisting of ECM and isotropic cytoskeletal structures) and contractile active fibers (e.g., actomyosin bundles).
The background material is modeled as an isotropic linear elastic medium, while the orientation and strength of the anisotropic contractile fibers is quantified by the nematic field, $\mathbf{Q}(\mathbf{x},t)$, a symmetric and traceless rank-2 tensor.
Feedback between stresses and fiber orientation controls the mechanical and structural behavior of the tissue.
Geometry plays a key role through the boundary conditions required when solving the force balance equation. 

\subsection{Mechanics}
Deformations of the background elastic material are described by the linear strain tensor ${\bm\varepsilon\equiv\tfrac{1}{2}[ \bm\nabla\mathbf{u}+(\bm\nabla\mathbf{u})^\mathsf{T}}]$ defined in terms of the displacement field $\mathbf{u}(\mathbf{x},t)$.
Assuming isotropic elastic properties, the stress-strain relation is simply
\begin{equation}
    \bm{\sigma}_\mathrm{e} = 2\mu \tilde{\bm\varepsilon} + B\,\mathrm{tr}[\bm\varepsilon]\mathbf{1}\,,
\end{equation}
where $\Tilde{\bm\varepsilon}$ denotes the deviatoric strain in the tissue and $\mathbf{1}$ the identity tensor.
The elastic response is fully described by
the bulk $B$ and the shear modulus $\mu$.

The active fibers exert a contractile fiber stress in the direction of their alignment, 
\begin{align}
\tilde{\bm\sigma}_\mathrm{f}=\alpha \mathbf{Q}\,,
\end{align}

where  $\alpha>0$ is the contractility and $\mathbf{Q}(\mathbf{x},t)$ is a nematic tensor field describing the local alignment of the fibers. This fiber stress is purely deviatoric. Contractile units such as myosin minifilaments that pull only along the fiber axis additionally generate an isotropic contribution to the active stress, which we refer to as bulk contractility. We neglect it in Secs.~\ref{sec:steady-states}--\ref{sec:quasi-1D-domain-wall}
 and discuss its effect in Sec.~\ref{sec:bulk-contractility}. 

Fiber remodeling, which are changes in $\mathbf{Q}(\mathbf{x},t)$, requires biochemical cellular processes and is generally slow compared to the mechanical relaxation of the tissue.
Therefore, we assume the tissue to be in mechanical equilibrium. The total stress ${\bm\sigma=\bm\sigma_\mathrm{e}+\bm\sigma_\mathrm{f}}$ then fulfills the force-balance condition
\begin{equation}\label{eq:force-balance}
    \bm\nabla\cdot\bm\sigma = 0\,.
\end{equation}
This equation needs to be solved with boundary conditions appropriate to the physical situation and geometry of interest.

\subsection{Mechanical feedback in fiber remodeling}
The central goal is to understand the role of mechanical feedback through which the tissue cells respond to the local stress environment and remodel the contractile fibers.
The dynamics of fiber remodeling is governed by the equation 

\begin{equation}
 \partial_t \mathbf{Q}=-\left(a + b\, \mathrm{tr}[\mathbf{Q}^2]\right)\mathbf{Q} + K \nabla^2 \mathbf{Q}+ \chi_\mathrm{f} \tilde{\bm\sigma}_\mathrm{f} + \chi_\mathrm{e} \tilde{\bm\sigma}_\mathrm{e}\,.
 \label{eq:fiber-dyn}
\end{equation}

We choose the constants ${a,b}>0$ to be positive to model that, in the absence of applied mechanical load, the fibers decay and disorder.  The gradient term describes liquid crystal elasticity, with $K$ the Frank elastic constant under the simplification of isotropic nematic elasticity. We have set the rotational viscosity of the nematic to unity. The coefficients $a$, $b$, and $K$ here are not meant to describe a thermodynamic free energy; instead, they parameterize the minimal relaxational dynamics. Fiber alignment decays at a rate $a$, its growth saturates through $b$, and distortions of the alignment are penalized by $K$. 
The last two terms in Eq.~\eqref{eq:fiber-dyn} capture the mechanical feedback on the fiber remodeling, with ${\chi_\mathrm{e,f}>0}$.  Both terms couple to the deviatoric part of the stress alone: an isotropic stress does not distinguish any direction in the plane and therefore does not contribute to fiber remodeling. The positive sign of the coefficients ${\chi_\mathrm{e,f}>0}$ highlights the central assumption of the model:  fibers align parallel to the direction along which tension is sustained.

Cells achieve such a response through several distinct mechanisms that act on different scales. First, tension stabilizes preexisting contractile structures. The actomyosin complex behaves as a catch-bond, whose lifetime increases with load \cite{guoMechanicsActomyosinBonds2006}. Tension on the actin filaments, moreover, suppresses the binding of cofilin and thereby protects the filaments from severing \cite{hayakawaActinFilamentsFunction2011}. Additionally, myosin and crosslinkers such as $\alpha$-actinin accumulate in regions of increased mechanical stress \cite{schiffhauerMechanoaccumulativeElementsMammalian2016}. 
Second, focal adhesions grow under applied force through the force-induced unfolding of talin and the recruitment of vinculin 
~\cite{delrioStretchingSingleTalin2009}. Adherens junctions are reinforced through the tension-dependent unfolding of $\alpha$-catenin~\cite{yonemuraACateninTensionTransducer2010} and the catch bond like binding of the cadherin-catenin couple to actin filaments ~\cite{buckleyMinimalCadherinCateninComplex2014}. 
Third, on longer time scales, sustained tension promotes the assembly of new contractile material.
Mechanosensitive ion channels convert membrane tension into intracellular signals~\cite{costePiezo1Piezo2Are2010}, and mechanosensitive transcriptional programs such as YAP/TAZ translate sustained mechanical load into changes of the cytoskeletal program of the cell~\cite{dupontRoleYAPTAZ2011}.
At the scale of whole bundles, this results in the load-dependent ordering of sarcomeres into myofibrils~\cite{kolley-kochelPhysicalBasisSarcomere2025}.

These mechanisms differ in molecular detail, but overall act as a positive feedback: contractile fibers are reinforced along the direction in which tension is sustained. This phenomenon is clearly exemplified by the cell-scale maturation of stress and muscle fibers under tension \cite{kolley-kochelPhysicalBasisSarcomere2025}.
We describe this feedback in minimal terms by introducing a simplified linear coupling (the terms proportional to $\chi_{\mathrm{e,f}}$) between fiber alignment and local deviatoric stresses within a continuum description. Since the background is linearly elastic, coupling to the deviatoric elastic stress is equivalent to coupling to the deviatoric strain, up to a factor of the shear modulus. 
We keep the two coupling coefficients distinct because the stress acting on an individual fiber and on its anchorages is not the coarse-grained stress of the tissue. A fiber is loaded both by the surrounding material and by its own contraction together with that of its immediate neighbors, so that the local stress it experiences differs from the average stress, in analogy to the cavity field of a polarizable medium \cite{friedrich2011nematic}. For $\chi_\mathrm{f} = \chi_\mathrm{e}$, the two terms combine to form the total deviatoric stress.

\begin{figure}[t]
    \centering
    \includegraphics[width=\columnwidth]{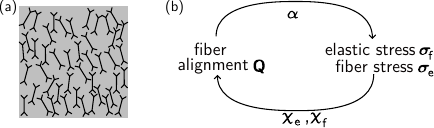}
    \caption{(a) Schematic describing the tissue as an elastic solid (in gray) with anisotropic contractile fibers (black; in the following figures as indicated as gray lines) embedded in the solid (gray). (b) Mechanical feedback between the fiber alignment and stress in the tissue. Fibers create contractile stress ${\alpha Q}$ which modifies both elastic ${\tilde{\bm\sigma}_\mathrm{e}}$ and fiber stress ${\tilde{\bm\sigma}_\mathrm{f}}$. The stresses in turn lead to fiber remodeling with coefficients $\chi_\mathrm{e}, \chi_\mathrm{f}$ 
    }
    \label{fig:model-schematic}
\end{figure}

The model combines three ingredients: an elastic background, contractile fibers, and the remodeling dynamics in which fibers reorient according to the local deviatoric stress. The mechanical boundary conditions enter through the force balance, and, as we show in the following sections,  dictate which fiber configurations can be maintained by the tissue [Fig.~\ref{fig:model-schematic}]. 

By substituting the expression for the fiber stress, the equation for the nematic tensor describing fiber alignment and orientation can be written as
\begin{subequations}
\begin{align}
    \partial_t \mathbf{Q} 
    = (\chi_\mathrm{f}\alpha - a) \mathbf{Q} - b \mathrm{Tr}[\mathbf{Q}^2]\mathbf{Q} + K \nabla^2 \mathbf{Q} + \chi_\mathrm{e} \tilde{\bm\sigma}_\mathrm{e}\,.\label{eq:fiber-feedback}
\end{align}
\end{subequations}
This equation will be solved below using no-flux boundary conditions for $\mathbf{Q}$, i.e., ${\mathbf{n}\cdot\bm\nabla\mathbf{Q}=0}$, with $\mathbf{n}$ the boundary normal, so that no fiber alignment is imposed at the boundary of the tissue. 
Equation~\eqref{eq:fiber-feedback}  shows that contractility alone can change the sign of the coefficient of the term linear in $\mathbf{Q}$  and build up nematic order when $\chi_f\alpha>a$. 
In the next section, we discuss the spontaneous alignment in detail, showing that the coupling to elastic stress modifies the transition by self-amplifying fiber alignment.

\section{Spontaneous alignment due to stress feedback}
\label{sec:steady-states}

To understand the structural patterning and mechanical response of the active elastic tissue with stress feedback, we first analyze the steady-state fiber configurations of a homogeneous tissue by solving Eq.~\eqref{eq:fiber-feedback}  for the homogeneous steady states with${\partial_t \mathbf{Q}=0}$ and $K \nabla^2 \mathbf{Q}=0$.
We find that stress feedback induces fiber self-alignment, and the coupling to elastic stresses orients the fibers with respect to the mechanical boundary conditions.
In this section we assume the tissue to be spatially homogeneous. Therefore, the fiber alignment $\mathbf{Q}$ does not depend on spatial position, and the strain $\mathbf{\varepsilon}$ and the stress $\mathbf{\sigma}$ are uniform and set by the boundary conditions.
We will denote quantities evaluated in the homogeneous steady states with an asterisk.

\subsection{Fiber-stress feedback: Spontaneous alignment }
\label{sec:chie0}

\begin{figure}[t]
    \centering
    \includegraphics[width=\columnwidth]{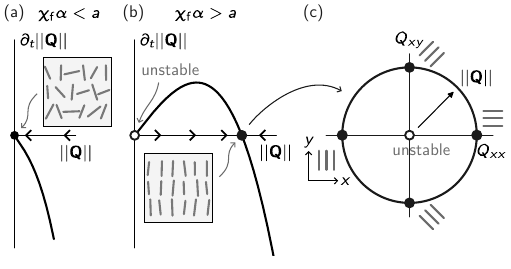}
    \caption{Fiber alignment caused by fiber-stress feedback.
    (a) In a homogeneous tissue, and for ${\chi_\mathrm{e}=0}$, fibers evolve (flow depicted by arrows) toward a disordered state with $||\mathbf{Q}||=0$ if ${\chi_\mathrm{f}\alpha<a}$, but spontaneously order for ${\chi_\mathrm{f}\alpha>a}$, because the disordered state becomes unstable (b).
    (c) Due to the rotational symmetry of the fiber dynamics for ${\chi_\mathrm{e}=0}$, the fiber direction is selected spontaneously
    The result is a degenerate ring of steady states in the $(Q_{xx},Q_{xy})$ plane corresponding to different fiber directions, shown for four selected steady states marked by black dots.
    }
    \label{fig:intro-selfAlignment}
\end{figure}

In the absence of feedback from the elastic stress, i.e., setting ${\chi_e=0}$, the steady-state fiber alignment $\mathbf{Q}^*$ is given by the solution of

\begin{align}
    (\chi_\mathrm{f}\alpha - a- b \mathrm{Tr}[\mathbf{Q}^{*2}]) \mathbf{Q}^*  =0\,.
\end{align}
It is evident that feedback from the contractile fiber stress is sufficient to drive alignment. For weak fiber-stress feedback (${\chi_\mathrm{f}\alpha<a}$) the stable solution is ${\mathbf{Q}^*=0}$ [Fig.~\ref{fig:intro-selfAlignment}(a)]. 
For ${\chi_\mathrm{f}\alpha>a}$, the ${\mathbf{Q}^*=0}$ state is unstable and fibers spontaneously align into a state with  ${||\mathbf{Q}^*||=\sqrt{(\chi_\mathrm{f}\alpha-a)/2b}>0}$, where ${\|\mathbf{Q}\|\equiv\sqrt{\mathrm{tr}[\mathbf{Q}^{2}]/2}}$ [Fig.~\ref{fig:intro-selfAlignment}(b)].  
 Fiber alignment and growth are induced by fiber-stress feedback: aligned fibers exert a contractile stress along their axis, which puts ``neighboring'' fibers under tension.
This promotes further alignment of fibers along the same direction once the feedback strength ${\chi_\mathrm{f}\alpha}$ exceeds the spontaneous decay rate $a$ of the fibers. 

Importantly, for ${\chi_\mathrm{e}=0}$, the fiber dynamics does not depend on the external load, and the direction of alignment is chosen spontaneously. The system remains rotationally symmetric, and the aligned state is therefore degenerate in direction.
These degenerate states form a circle of radius ${\|\mathbf{Q}\|}$ in the ${(Q_{xx}, Q_{xy})}$ plane [see Fig~\ref{fig:intro-selfAlignment} (c)], which fully parametrizes the nematic tensor since it is symmetric and traceless, hence $Q{_{yx}=Q_{xy}}$ and ${Q_{yy}=-Q_{xx}}$. 

\subsection{Elastic stress feedback: two orthogonal states}
\label{sec:chie}

\begin{figure}[t]
    \centering
    \includegraphics[width=\columnwidth]{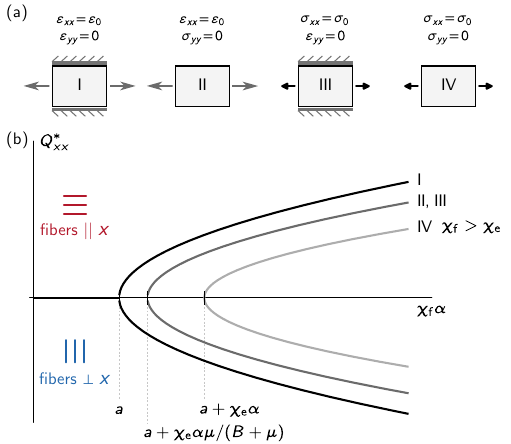}
    \caption{Alignment transition under different boundary conditions.   (a) Sketches of the four different mechanical boundary conditions.  (b) The onset of self-alignment under these boundary conditions differs. The steady-state alignment ${Q^*_{xx}}$ is calculated as the solution of Eq.~\eqref{eq:flow-general}. Case IV only allows self-alignment if ${\chi_\mathrm{f}>\chi_\mathrm{e}}$.}
    \label{fig:alignment-transition}
\end{figure}
We now consider the effect of coupling to the elastic stress, setting $\chi_e \neq0$.
We note that the elastic stress is determined by both the active contractility of the fibers embedded in the elastic material and any external load through the solution of the force balance equation with suitable  boundary conditions.
Accordingly, the elastic-stress feedback has two effects.
First, the elastic stress created by the contractility of the fibers will lead to an indirect fiber-feedback term that modifies the self-alignment transition.
Second, the mechanical boundary conditions will generically break the rotational symmetry of the system and select the possible fiber directions as parallel and perpendicular to the externally applied load.
In Sec. \ref{sec:lsa}, we show that these two directions are the only stable homogeneous states: states with fibers tilted relative to the load are always unstable.

To make both effects explicit, we consider a rectangular tissue patch loaded along $x$, and compare different combinations of strain and stress boundary conditions.
We first look for steady states in which the fibers are aligned with the axes of the load, that is, parallel or perpendicular to $x$, so that $Q^{*}_{xy}=0$. In Sec. \ref{sec:lsa}, we show by a linear stability analysis that these states are indeed stable for all boundary conditions considered. In this case, one has ${\mathrm{tr}[\mathbf{Q}^{*2}]=2(Q^{*}_{xx})^{2}}$, reducing Eq.~\eqref{eq:fiber-feedback} to
\begin{equation}\label{eq:uniax-strain-hss}
    (\chi_\mathrm{f}\alpha - a) Q_{xx}^{*} - 2b {Q^{*}_{xx}}^3 + 2\chi_\mathrm{e}\mu \tilde\varepsilon_{xx}^{*}=0\,,
\end{equation}
where ${\tilde\varepsilon^{*}_{xx}=(\varepsilon^{*}_{xx}-\varepsilon^{*}_{yy})/2}$ is the deviatoric strain along $x$. 
 The circle of steady states obtained in Sec \ref{sec:chie0} reduces to two orthogonal aligned states, which we denote by ${Q^{+}_{xx}>0}$, fibers parallel to the load, and ${Q^{-}_{xx}<0}$, fibers perpendicular to the load.
Next, we use the force-balance condition Eq.~\eqref{eq:force-balance} to solve for the deviatoric strain in terms of the nematic alignment $Q_{xx}^*$, use this result to close Eq.~\eqref{eq:uniax-strain-hss}, and calculate the fixed points. 
   
Along each direction, one can control one of two quantities (or a combination): the displacement of the boundary, which for a homogeneous patch fixes the corresponding component of the strain, or the force per unit length applied to the boundary (the traction), which fixes the corresponding component of the total stress.
To study the effect of different boundary conditions, we load the patch along $x$, either by imposing a strain $\varepsilon_{0}$ that stretches the patch or by pulling it with a stress $\sigma_{0}$ (both positive for extension), and either clamp the transverse direction ($\varepsilon_{yy}=0$), or leave it free ($\sigma_{yy}=0$).
This gives four cases [Fig.~\ref{fig:alignment-transition}(a)]:
\begin{itemize}
    \item[I] \emph{Clamped stretching}, $\varepsilon_{xx}=\varepsilon_{0}$ and $\varepsilon_{yy}=0$: the patch is stretched while its width is held fixed.
    \item[II] \emph{Free stretching}, $\varepsilon_{xx}=\varepsilon_{0}$ and $\sigma_{yy}=0$: the patch is stretched in $x$, and its lateral edges are free, allowing it to change its width in the $y$ direction.
    \item[III] \emph{Clamped loading}, $\sigma_{xx}=\sigma_{0}$ and $\varepsilon_{yy}=0$: pulling with a fixed force on the patch while its width is held fixed. 
    \item[IV] \emph{Free loading}, $\sigma_{xx}=\sigma_{0}$ and $\sigma_{yy}=0$: pulling with a fixed force while its width in the $y$ direction is free to change.
\end{itemize}
Here, a clamped direction means that the corresponding normal displacement of the boundary is prescribed, and a free direction means that the boundary carries no traction at all. 
None of these boundary conditions prescribe an off-diagonal strain $\varepsilon^{*}_{xy}$ or stress$\sigma^{*}_{xy}$.
Under the assumption ${Q_{xy}^*=0}$, the fibers do not exert off-diagonal stress, and the off-diagonal strain ${\varepsilon^{*}_{xy}=0}$ always vanishes.
In contrast, the deviatoric-strain $\tilde\varepsilon^{*}_{xx}$, which enters the fiber dynamics in Eq.~\eqref{eq:fiber-dyn} depends on the boundary conditions.  The two normal stresses are given by
\begin{subequations}\label{eq:stress-components}
\begin{align}    \sigma_{xx}&=2\mu\,\tilde\varepsilon_{xx}+B\,(\varepsilon_{xx}+\varepsilon_{yy})+\alpha Q_{xx}\,,\label{eq:sigma-xx}\\   \sigma_{yy}&=-2\mu\,\tilde\varepsilon_{xx}+B\,(\varepsilon_{xx}+\varepsilon_{yy})-\alpha Q_{xx}\,.\label{eq:sigma-yy}
\end{align}
\end{subequations}
For the chosen boundary conditions, the force-balance equation Eq.~\eqref{eq:force-balance} implies that the stresses are constant. We now determine the deviatoric strain for each of the four cases. 

\textbf{Case I: clamped stretching.} Both strain components are prescribed, $\varepsilon_{xx}=\varepsilon_{0}$ and $\varepsilon_{yy}=0$, so the deviatoric strain follows directly from its definition as
\begin{equation}\label{eq:deveps-I}
    \tilde\varepsilon^{*}_{xx}=\frac{\varepsilon_{0}}{2}\,.
\end{equation}
It does not depend on the fiber alignment because the patch cannot deform. Any stresses exerted by the fibers or caused by elastic deformations are supported by the clamped boundaries.

\textbf{Case II: free stretching.} Here, only $\varepsilon_{xx}=\varepsilon_{0}$ is prescribed while the transverse strain $\varepsilon^{*}_{yy}$ follows from the force-balance equation under the condition that the lateral edges are free, ${\sigma_{yy}=0}$.
Thus, inserting $\tilde\varepsilon^{*}_{xx}=(\varepsilon_{0}-\varepsilon^{*}_{yy})/2$ into Eq.~\eqref{eq:sigma-yy} we can solve 

 for the transverse strain, with the result
    \begin{equation}\label{eq:caseII-eyy}       \varepsilon^{*}_{yy}=\frac{\alpha Q^{*}_{xx}-(B-\mu)\,\varepsilon_{0}}{B+\mu}\,.
    \end{equation}
Inserting Eq.~\eqref{eq:caseII-eyy} into the definition of the deviatoric strain, we obtain
    \begin{equation}\label{eq:deveps-II}
        \tilde\varepsilon^{*}_{xx}=\frac{\varepsilon_{0}-\varepsilon^{*}_{yy}}{2}
        =\frac{B}{B+\mu}\,\varepsilon_{0}-\frac{\alpha}{2(B+\mu)}\,Q^{*}_{xx}\,.
    \end{equation}
The first term is the deviatoric strain of a passive tissue stretched by $\varepsilon_{0}$. The second term is generated by the fibers. Its negative sign reflects the tissue response to the contractile fibers. Because the patch is free to deform transversally, the fiber stress increases $\varepsilon^{*}_{yy}$ and thereby reduces $\tilde\varepsilon^{*}_{xx}$ accordingly. As the deviatoric strain feeds back on the fiber alignment, this term represents an indirect fiber feedback. 

\textbf{Cases III and IV: Clamped and free loading.} A similar procedure gives the deviatoric strain for the remaining cases, in which a load (hence a stress) is applied.  In case III, $\varepsilon_{yy}=0$ is prescribed and $\varepsilon_{xx}=2\tilde{\varepsilon}_{xx}$. Eq.~\eqref{eq:sigma-xx} with $\sigma_{xx}=\sigma_{0}$ gives $\sigma_{0}=(B+\mu)\,\varepsilon^{*}_{xx}+\alpha Q^{*}_{xx}$. In case IV, no strain is prescribed, and subtracting Eq.~\eqref{eq:sigma-yy} from Eq.~\eqref{eq:sigma-xx} gives $\sigma_{0}=4\mu\,\tilde\varepsilon^{*}_{xx}+2\alpha Q^{*}_{xx}$. Solving these relations for the deviatoric strain, one obtains,
\begin{subequations}\label{eq:deveps-cases}
\begin{align}    \text{III:}\quad\tilde\varepsilon^{*}_{xx}&=\frac{\sigma_{0}}{2(B+\mu)}-\frac{\alpha}{2(B+\mu)}\,Q^{*}_{xx}\,,\label{eq:deveps-III}\\  \text{IV:}\quad\tilde\varepsilon^{*}_{xx}&=\frac{\sigma_{0}}{4\mu}-\frac{\alpha}{2\mu}\,Q^{*}_{xx}\,.\label{eq:deveps-IV}
\end{align}
\end{subequations}
Both Cases II and III carry the same fiber-induced coefficient, $\tfrac{\alpha}{2(B+\mu)}$, because in both the patch is free to deform along one direction only, whereas Case IV, with both directions free to deform, gives the larger coefficient $\tfrac{\alpha}{2\mu}$.  

In each of the four cases, the deviatoric strain consists of a term set by the applied load and a term proportional to the fiber alignment. 
Inserting the expression for the deviatoric strain into the steady-state equation for the fiber alignment, Eq.~\eqref{eq:uniax-strain-hss}, this can be rewritten in the form,
\begin{equation}\label{eq:flow-general}
    f(Q^{*}_{xx})\equiv\tilde a \,Q^{*}_{xx}-2b\,{Q^{*}_{xx}}^{3}+h=0\,
\end{equation}
where $\tilde a \equiv\chi_\mathrm{f}\alpha-a_\mathrm{eff}$. The effective fiber decay rate $a_\mathrm{eff}$ and the biasing field $h$ are given  by 
\begin{subequations}\label{eq:aeff-h-values}
\begin{align}
    a_\mathrm{eff}&=
    \begin{cases}
        a\,, & \text{case I},\\[2pt]
        a+\dfrac{\mu}{B+\mu}\,\chi_\mathrm{e}\alpha\,, & \text{cases II, III},\\[6pt]
        a+\chi_\mathrm{e}\alpha\,, & \text{case IV},
    \end{cases}\label{eq:aeff-values}\\[6pt]
    h&=
    \begin{cases}
        \chi_\mathrm{e}\mu\,\varepsilon_{0}\,, & \text{case I},\\[4pt]
        \dfrac{2B}{B+\mu}\,\chi_\mathrm{e}\mu\,\varepsilon_{0}\,, & \text{case II},\\[6pt]
        \dfrac{\mu}{B+\mu}\,\chi_\mathrm{e}\,\sigma_{0}\,, & \text{case III},\\[6pt]
        \tfrac{1}{2}\,\chi_\mathrm{e}\,\sigma_{0}\,, & \text{case IV}.
    \end{cases}\label{eq:h-values}
\end{align}
\end{subequations}

\begin{figure}
    \centering
    \includegraphics[width=\columnwidth]{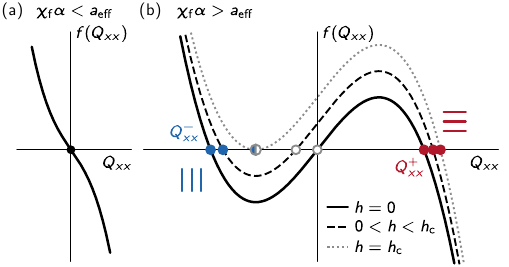}
    \caption{Homogeneous fiber steady states are given by the zeros of ${f(Q_{xx})}$. Filled circles mark the stable steady states, with red indicating alignment along the load direction $x$ and blue alignment along $y$. Open circles mark the unstable state between them.
    }
    \label{fig:flow}
\end{figure}

Without load (${h=0}$), Eq.~\eqref{eq:flow-general} has the same form as in the absence of elastic-stress feedback, $\chi_\mathrm{e}=0$, treated in Sec. \ref{sec:chie0} , with $a$ replaced by $a_\mathrm{eff}$: aligned states, $Q^{\pm}_{xx}=\pm\sqrt{\tilde a/2b}$, exist for ${\chi_\mathrm{f}\alpha>a_\mathrm{eff}}\geq a$ [see Fig.~\ref{fig:alignment-transition}(b)].
The transition is shifted to higher values of contractility because the elastic stress induced by the contractile fibers counteracts the fiber stress and is oriented opposite to the fiber stress.
or ${h=0}$, both states $Q^{+}_{xx}$ (fibers aligned along the loading direction $x$) and $Q^{-}_{xx}$ (fibers aligned along the transverse direction $y$) are stable, and they are related by  ${Q^-_{xx}\to-Q^+_{xx}}$ due to the symmetry of Eq.~\eqref{eq:flow-general}. The biasing term $h$ breaks this symmetry  and, for ${h>0}$, favors fibers along the load [Fig.~\ref{fig:flow}(b)].
 
The bias $h$ is linear in the applied load. For $\tilde a>0$ and small $h$, Eq.~\eqref{eq:flow-general} has three roots, which to linear order in $h$ read
    \begin{equation}\label{eq:Qpm}
        Q^{\pm}_{xx}\simeq\pm\sqrt{\frac{\tilde a}{2b}}+\frac{h}{2\tilde a}\,,\qquad
        Q^{0}_{xx}\simeq-\frac{h}{\tilde a}\,.
    \end{equation}
In the state $Q^{+}_{xx}>0$, the fibers are aligned parallel to the load. 
In the state $Q^{-}_{xx}<0$ they are aligned perpendicular to it. 
The intermediate root $Q^{0}_{xx}$ is unstable: an initial alignment above it evolves toward the parallel state, and an initial alignment below it towards the perpendicular state [Fig.~\ref{fig:flow}].

\textbf{Loss of the perpendicular state under strong load.} As $h$ increases, $Q^{-}_{xx}$ and $Q^{0}_{xx}$ approach each other and annihilate in a saddle-node bifurcation (see Fig.~\ref{fig:flow}), at which $f$ and its derivative $f'$ vanish simultaneously. This occurs at
\begin{equation}\label{eq:hss-saddle-node}
    |h|=h_\mathrm{c}=\frac{2\tilde a}{3}\sqrt{\frac{\tilde a}{6b}}\,.
\end{equation}
For $|h|>h_\mathrm{c}$ only the root $Q^{+}_{xx}$ remains [Fig.~\ref{fig:flow}].

\begin{figure}
    \centering
    \includegraphics[width=\columnwidth]{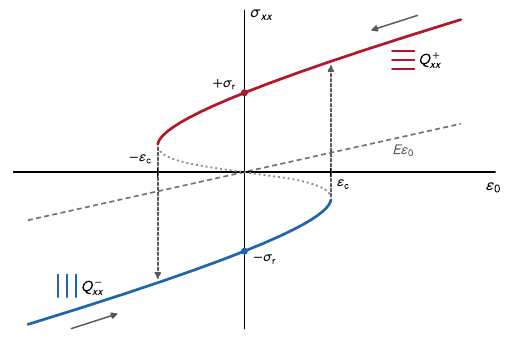}
    \caption{Stress--strain relation of a patch under free stretching (case II) [Eq.~\eqref{eq:response-cases}]. Red: fibers parallel to the load ($Q^{+}_{xx}$); blue: fibers perpendicular to the load ($Q^{-}_{xx}$); dashed: passive tissue, $\sigma_{xx}=E\varepsilon_{0}$. The dotted portion of the line is unstable. At zero strain, the two branches carry the residual stresses $\pm\sigma_\mathrm{r}$
    }
    \label{fig:stress-strain}
\end{figure}

\textbf{Mechanical bistability.} Because the fibers yield a contribution $\alpha\mathbf{Q}$ to the total stress, the two states differ not only in the direction of fiber alignment, but also in their mechanical response.
The stress-strain relation along the loading direction for each of the cases follows from Eqs.~\eqref{eq:stress-components}, \eqref{eq:deveps-II}, and~\eqref{eq:deveps-cases} as 
\begin{equation}\label{eq:response-cases}
\begin{aligned}
    \text{I:}&\quad \sigma_{xx}=(B+\mu)\,\varepsilon_{0}+\alpha Q^{*}_{xx}\,,\\
    \text{II:}&\quad \sigma_{xx}=E\,\varepsilon_{0}+\frac{2\alpha B}{B+\mu}\,Q^{*}_{xx}\,,\\
    \text{III:}&\quad \varepsilon^{*}_{xx}=\frac{\sigma_{0}-\alpha Q^{*}_{xx}}{B+\mu}\,,\\
    \text{IV:}&\quad \varepsilon^{*}_{xx}=\frac{\sigma_{0}}{E}-\frac{\alpha}{2\mu}\,Q^{*}_{xx}\,,
\end{aligned}
\end{equation}
where ${E\equiv4\mu B/(B+\mu)}$ is the two-dimensional Young modulus of the passive background. The aligned states ${Q_{xx}^*=Q^{\pm}_{xx}}$ themselves depend on the load through $h$ [Eq.~\eqref{eq:flow-general}], so that Eqs.~\eqref{eq:response-cases} have to be evaluated with the corresponding solution for each state. Each of the two states thus defines its own branch of the mechanical response, and both branches are continuous wherever the respective state exists. For small loads, Eq.~\eqref{eq:Qpm} gives, for instance in case II, $\sigma_{xx}\simeq\pm\sigma_\mathrm{r}+E_\mathrm{eff}\,\varepsilon_{0}$, with the residual stress $\sigma_\mathrm{r}=2\alpha B\sqrt{\tilde a/2b}/(B+\mu)$ and $E_\mathrm{eff}=E+2\chi_\mathrm{e}\alpha\mu B^{2}/[\tilde a(B+\mu)^{2}]$ [Fig.~\ref{fig:stress-strain}]. Whenever the parallel and the perpendicular states are both stable, the same applied load can induce two different values of the stress and strain [cf.\ Eq.~\eqref{eq:response-cases}]. As we show in the next section, both states are indeed stable under suitable boundary conditions, and the resulting bistability leads to a hysteretic response of the tissue, which we discuss in more detail below.

\section{Linear stability}
\label{sec:lsa}

\begin{figure}
    \centering
    \includegraphics[width=\columnwidth]{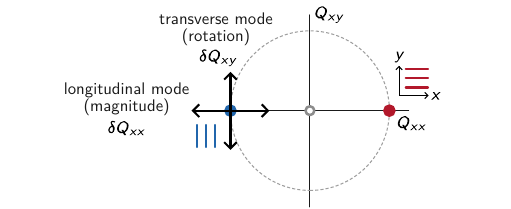}
    \caption{Linear stability analysis of the perpendicular fiber alignment.
    The fiber dynamics Eq.~\eqref{eq:fiber-feedback} is linearized in deviations of the fiber alignment ${\delta Q_{xx},\delta Q_{xy}}$ from the steady state of parallel (${Q_{xx}>0}$ , shown in red ) or perpendicular (${Q_{xx}<0}$, shown in blue) fiber alignment.
    Perturbations parallel to the steady-state fiber alignment (in the ${(Q_{xx}, Q_{xy})}$ plane) alter the magnitude of fiber alignment, while transverse perturbations rotate the fiber direction.
    }
    \label{fig:LSA-setup}
\end{figure}

In the Sec.~\ref {sec:steady-states}, we discussed the self-alignment transition of the fibers under external load.
We showed that self-alignment results in steady states in which the fibers are parallel or perpendicular to the load.
In this section, we analyze whether these two states are stable against small perturbations by performing a linear stability analysis (see Fig.~\ref{fig:LSA-setup}).
This analysis examines both the stability to perturbations that change the degree of alignment while keeping the direction fixed (longitudinal perturbations), as well as perturbations that rotate the fibers away from their original alignment direction (transverse perturbations) [Fig.~\ref{fig:LSA-setup}].
To study transverse perturbations, we additionally allow for shear deformations of the tissue that are induced by the rotating fibers.
We find that the fiber state $Q_{xx}^0$ is unstable to longitudinal perturbations [Sec.~\ref{sec:chie}].
Additionally, fibers parallel to the load are always stable, whereas fibers perpendicular to the load are stable only if the boundary conditions allow for sufficiently strong shear deformations of the tissue.
This condition can be fulfilled for free stretching and clamped loading, resulting in a structurally and mechanically bistable response of the tissue.

\textbf{Perturbation ansatz.} We write the fiber alignment and the deviatoric strain as the steady state plus small perturbations
\begin{subequations}\label{eq:lsa-ansatz}
    \begin{align}
        Q_{xx} &= Q_{xx}^*+\delta Q_{xx}\,, & Q_{xy} &= \delta Q_{xy}\,,\\
        \tilde\varepsilon_{xx} &= \tilde\varepsilon^*_{xx}+\delta\tilde\varepsilon_{xx}\,, & \tilde\varepsilon_{xy} &= \delta\tilde\varepsilon_{xy}\,,
    \end{align}
\end{subequations}
where ${Q^{*}_{xx}}$ is either the parallel state ${Q^{+}_{xx}}$ or the perpendicular state ${Q^{-}_{xx}}$, and ${\tilde\varepsilon^{*}_{xx}}$ is the corresponding deviatoric strain [cf.\ Eqs.~\eqref{eq:deveps-I}, \eqref{eq:deveps-II}, and~\eqref{eq:deveps-cases}]. The perturbation ${\delta Q_{xx}}$ changes the degree of alignment at fixed direction (\emph{amplitude/longitudinal mode}), and ${\delta Q_{xy}}$ rotates the fibers away from the loading axis (\emph{rotation/transverse mode}) (see Fig.~\ref{fig:LSA-setup}).  
Inserting the perturbation ansatz Eq.~\eqref{eq:lsa-ansatz} into the fiber dynamics Eq.~\eqref{eq:fiber-feedback}, and linearizing, we find
\begin{subequations}\label{eq:lsa-general}
\begin{align}
    \partial_t \delta Q_{xx} &\approx \big[\chi_\mathrm{f}\alpha-a-6b\,{Q^{*}_{xx}}^{2}\big]\delta Q_{xx} +2\chi_\mathrm{e}\mu\,\delta\tilde\varepsilon_{xx}\,,\label{eq:lsa-general-long}\\
    \partial_t \delta Q_{xy} &\approx \big[\chi_\mathrm{f}\alpha-a-2b\,{Q^{*}_{xx}}^{2}\big]\delta Q_{xy} +2\chi_\mathrm{e}\mu\,\delta\tilde\varepsilon_{xy}\,.\label{eq:lsa-general-trans}
\end{align}
\end{subequations}

The strain perturbations in Eqs.~\eqref{eq:lsa-general} depend on the fiber perturbations through the force-balance condition, which the tissue satisfies at all times because the fiber remodeling is assumed to be slow compared to the mechanical relaxation of the tissue. To obtain the strain perturbations, we insert the perturbation ansatz Eq.~\eqref{eq:lsa-ansatz} into the force-balance condition and solve for the deviatoric strain, as in Sec. \ref{sec:chie}. The diagonal stress components in the frame of the load, Eq.~\eqref{eq:stress-components}, do not involve $Q_{xy}$, so that one obtains the same expression given in Eqs.~\eqref{eq:deveps-I}, \eqref{eq:deveps-II}, and~\eqref{eq:deveps-cases}, but now evaluated at the perturbed alignment $Q_{xx}^*+\delta Q_{xx}$. Since the load, and hence the biasing field $h$, is held fixed during the perturbation, the steady-state relation gives, 
\begin{equation}\label{eq:lsa-amp-strain}
    2\chi_\mathrm{e}\mu\,\delta\tilde\varepsilon_{xx}=-(a_\mathrm{eff}-a)\,\delta Q_{xx}\,.
\end{equation}
In case I, the perturbation $\delta\varepsilon_{xx}$ vanishes because both normal strains are prescribed, and ${a_\mathrm{eff}=a}$.
The shear strain, in contrast, is not fixed by the diagonal stress components. A rotation of the fibers, in turn, creates an off-diagonal fiber stress $\alpha\,\delta Q_{xy}$, i.e., a shear stress, so that the total shear stress is $\delta\sigma_{xy}=2\mu\,\delta\tilde\varepsilon_{xy}+\alpha\,\delta Q_{xy}$. The tissue's response depends on the boundary conditions. In case I, the displacement is prescribed for both $u_x$ and $u_y$ along the whole boundary; therefore, the shear strain $\delta\tilde\varepsilon_{xy}=0$. In cases II, III, and IV, in contrast, at least one pair of edges is free and therefore carries no traction, while the remaining edges prescribe only the displacement normal to them. The tangential traction thus vanishes along the whole boundary in these three cases, implying that the total shear stress vanishes, $\sigma_{xy}=0$. Thus, the tissue shears until its elastic shear stress balances the fiber stress, that is, one has
\begin{equation}\label{eq:lsa-induced-shear}
    2\mu\,\delta\tilde\varepsilon_{xy}=-\alpha\,\delta Q_{xy}\,.
\end{equation}
Inserting the strain responses, Eqs.~\eqref{eq:lsa-amp-strain} and~\eqref{eq:lsa-induced-shear}, into the linearized dynamics, Eqs.~\eqref{eq:lsa-general}, shows that the dynamics of longitudinal and transverse perturbations decouple: the longitudinal strain responds only to $\delta Q_{xx}$ and the shear strain only to $\delta Q_{xy}$. We thus obtain ${\partial_t\delta Q_{xx}\approx\lambda_{\parallel}\,\delta Q_{xx}}$ and ${\partial_t\delta Q_{xy}\approx\lambda_{\perp}\,\delta Q_{xy}}$, with the growth rates
\begin{subequations}\label{eq:lsa-decoupled}
\begin{align}
    \lambda_{\parallel}&=\chi_\mathrm{f}\alpha-a_\mathrm{eff}-6b\,{Q^{*}_{xx}}^{2}\,,\label{eq:lsa-decoupled-long}\\
    \lambda_{\perp}&=\chi_\mathrm{f}\alpha-a-2b\,{Q^{*}_{xx}}^{2}-
    \begin{cases}
        0\,, & \text{shear fixed},\\
        \chi_\mathrm{e}\alpha\,, & \text{shear free}.
    \end{cases}\label{eq:lsa-decoupled-trans}
\end{align}
\end{subequations}

Note that the effective decay rate $a_\mathrm{eff}$ appears only in the longitudinal growth rate, because it accounts for the normal deformation that the fibers induce [Eq.~\eqref{eq:lsa-amp-strain}]. The transverse growth rate instead contains the bare decay rate $a$ together with the shear term, which accounts for the shear deformation induced by a rotation of the fibers [Eq.~\eqref{eq:lsa-induced-shear}]. We discuss the two modes in turn.

\begin{figure*}
    \centering
    \includegraphics[]{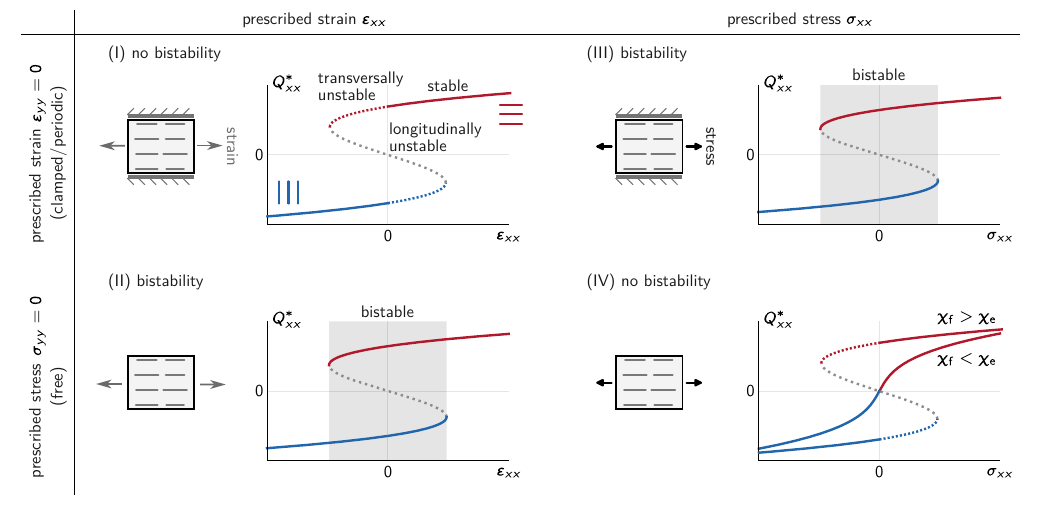}
  \caption{Bistability of a homogeneous tissue patch.
  Steady states $Q^{*}_{xx}$ are shown as a function of the applied load for the four boundary conditions of Fig.~\ref{fig:alignment-transition}(a).
  The fiber alignment $Q^{*}_{xx}$ is given by the solutions of Eq.~\eqref{eq:flow-general}.
  The rows of the table describe whether the transverse direction of the patch is clamped, ${\varepsilon_{yy}=0}$ (top), or free, ${\sigma_{yy}=0}$ (bottom).
  The columns distinguish prescribed strain $\varepsilon_{0}$ (left) or prescribed stress $\sigma_{0}$ (right) in the loading direction $x$. Stable fiber alignment states are labeled as solid lines, while grey dotted lines mark states that are unstable against amplitude perturbations ($Q^{0}_{xx}$).
  The colored dotted lines denote steady states unstable against fiber rotation.
  The range of bistability is shaded in grey and shown for the case that both alignment branches (red and blue) are stable over their full range of existence.
  In Case IV, several steady-states of fiber alignment exist only for ${\chi_\mathrm{f}>\chi_\mathrm{e}}$.}
    \label{fig:bistability}
\end{figure*}

\textbf{Amplitude mode.} The growth rate $\lambda_{\parallel}=\tilde a-6b\,(Q^{*}_{xx})^{2}$ is the slope of the function $f(Q_{xx})$, [Eq.~\eqref{eq:flow-general}], evaluated at the steady states. Graphically, $f$ crosses zero with a negative slope at values of $Q_{xx}$ corresponding to the two aligned states and with a positive slope at the intermediate state $Q^{0}_{xx}$ [Fig. \ref{fig:flow}]. Both aligned states are therefore stable against longitudinal perturbations for all four boundary conditions, and the perpendicular state becomes marginally stable only at the saddle node, where it merges with $Q^{0}_{xx}$ and vanishes. Whether the patch is bistable is therefore decided by the rotation mode.

\textbf{Rotation mode.}. Now, we focus on the stability of the rotation mode and explicitly evaluate the transverse growth rate.
Dividing the steady-state condition, Eq.~\eqref{eq:flow-general}, by ${Q^{*}_{xx}}$ gives ${2b\,{Q^{*}_{xx}}^{2}=\tilde a+h/Q^{*}_{xx}}$, so that the transverse growth rate, Eq.~\eqref{eq:lsa-decoupled-trans}, depends on the steady state only through the ratio ${h/Q^{*}_{xx}}$ of the biasing field to the alignment.
This ratio is positive for the parallel state, which the load strengthens, and negative for the perpendicular state, which the load weakens.
Using the effective decay rate ${a_\mathrm{eff}}$ given in Eqs.~\eqref{eq:aeff-h-values} and the shear conditions discussed above Eq~\eqref{eq:lsa-decoupled-trans}, one obtains
\begin{equation}\label{eq:lsa-rotation-cases}
\begin{aligned}
    \text{I, IV:}&\quad \lambda_{\perp}=-\frac{h}{Q^{*}_{xx}}\,,\\
    \text{II, III:}&\quad \lambda_{\perp}=-\frac{h}{Q^{*}_{xx}}-\frac{B}{B+\mu}\,\chi_\mathrm{e}\alpha\,.
\end{aligned}
\end{equation}
Consequently, the load stabilizes the parallel state ($Q_{xx}^*>0$) against rotations and destabilizes the perpendicular one ($Q_{xx}^*<0$).
The parallel fiber configuration is thus stable for all four boundary conditions [Fig.~\ref{fig:bistability}, cases I--IV]. 
The perpendicularly aligned state $Q_{xx}^-<0$ is unstable for boundary conditions I and IV, where the elastic stress feedback does not resist the rotation: any load makes the fibers rotate into the parallel state. 

For cases II and III, the elastic-stress feedback can resist the rotation, and $\lambda_{\perp}<0$ as long as
\begin{equation}\label{eq:rotation-stability}
    \frac{h}{|Q^{-}_{xx}|}<\frac{B}{B+\mu}\,\chi_\mathrm{e}\alpha\,.
\end{equation}
Since the left-hand side vanishes without load, the tissue is bistable under sufficiently weak loads. The perpendicular state exists only between its value without load and the saddle node, ${\sqrt{\tilde a/6b}<|Q^{-}_{xx}|<\sqrt{\tilde a/2b}}$ [Fig.~\ref{fig:flow}], so that ${h/|Q^{-}_{xx}|<\tfrac{2}{3}\tilde a}$. 
The perpendicular state is therefore stable over its entire range of existence, ${|h|<h_\mathrm{c}}$, whenever
\begin{equation}\label{eq:rotation-cond}
    \tilde a\leq\frac{3}{2}\,\frac{B}{B+\mu}\,\chi_\mathrm{e}\alpha\,,
\end{equation}
or equivalently $\chi_\mathrm{f}\alpha-a\leq\chi_\mathrm{e}\alpha\,(\mu+3B/2)/(B+\mu)$. The right-hand side of the latter is at least $\chi_\mathrm{e}\alpha$, while the left-hand side is smaller than $\chi_\mathrm{f}\alpha$. For comparable couplings, $\chi_\mathrm{f}\approx\chi_\mathrm{e}$, the condition is therefore satisfied. In this case, the perpendicular state does not lose its stability against rotations before it disappears at the saddle node, so that the tissue is bistable in the entire range of loads in which both states exist, ${|h|<h_\mathrm{c}}$ [see Fig.~\ref{fig:bistability} cases II and III].

\textbf{Mechanical bistability.}
The above discussion shows that the tissue is structurally bistable in cases~II and~III, that is, two orthogonal fiber configurations are stable under the same load. Because the fibers create contractile stress, this structural bistability implies a bistable mechanical response of the tissue: under free stretching (case II), the same applied strain results in two different stresses, whereas under clamped loading (case III), where the stress is prescribed, the same applied stress results in two different strains [cf.\ Eq.~\eqref{eq:response-cases} and Fig.~\ref{fig:bistability}, cases II and III]. In both cases, the mechanical response of the tissue depends on the loading history and exhibits hysteresis. 

To illustrate this behavior, consider the case of free stretching (Case II). As the patch is stretched, the strain increasingly favors the parallel fiber state.
At the critical stretch $\varepsilon_\mathrm{c}=(B+\mu)h_\mathrm{c}/(2B\chi_\mathrm{e}\mu)$, at which the biasing field reaches $h_\mathrm{c}$ [cf.\ Eqs.~\eqref{eq:h-values} and~\eqref{eq:hss-saddle-node}], the perpendicularly aligned steady state disappears, and the fibers always align parallel to the applied stretch. When the stretch is (adiabatically) released to zero, the fibers remain parallel: the fiber stress feedback sustains their alignment, since the fibers keep pulling on each other even though the biasing field vanishes. The patch can therefore keep the residual stress $\sigma_\mathrm{r}=2\alpha B\sqrt{\tilde a/2b}/(B+\mu)$ at zero stretch.
The fibers only flip into the perpendicular state when the patch is compressed beyond $-\varepsilon_\mathrm{c}$, where the parallel state disappears at the saddle node and only the perpendicularly aligned state remains, or when the parallel state becomes unstable, if the threshold Eq.~\eqref{eq:rotation-stability} is reached before. The stress-strain curve thus traces a hysteresis loop [Fig.~\ref{fig:stress-strain}]. In clamped loading (case III), the roles of stress and strain are exchanged: under the same prescribed force, the two states show different strains instead of different stresses. [Eq.~\eqref{eq:response-cases}].

\section{Effective potential}
\label{sec:potential}

Since the force-balance condition is linear in the fiber alignment, it can be solved not only for small perturbations around the load-aligned fiber states as in the previous section, but also for an arbitrary alignment.
The deviatoric strain along the load axis is then still given by Eqs.~\eqref{eq:deveps-I},~\eqref{eq:deveps-II}, and~\eqref{eq:deveps-cases} for the different mechanical boundary conditions. 
The shear strain can be calculated as in the linear stability analysis [cf.\ Eqs.~\eqref{eq:lsa-amp-strain},~\eqref{eq:lsa-induced-shear}] and is given by ${2\mu\,\tilde\varepsilon_{xy}=-\alpha Q_{xy}}$ if the tissue can shear and vanishes if it cannot. The steady-state alignment ${Q_{xx}^*}$ is, however, modified as described below.
Inserting both expressions for the deviatoric strain into the fiber dynamics, Eq.~\eqref{eq:fiber-feedback}, gives closed dynamical equations for the two components of the alignment tensor $\{Q_{xx}, Q_{xy}\}$.
This dynamics can be expressed as a gradient flow, 
\begin{equation}
    \partial_tQ_{xx}=-\frac{\partial V}{\partial Q_{xx}}\,, \qquad \partial_tQ_{xy}=-\frac{\partial V}{\partial Q_{xy}}\,,
\end{equation} 
with respect to the effective potential
\begin{align}\label{eq:potential}
    V=&-\frac{\tilde a}{2}\,\|\mathbf{Q}\|^{2}+\frac{b}{2}\,\|\mathbf{Q}\|^{4}-h\,Q_{xx}\nonumber\\
    &+\frac{Q_{xy}^{2}}{2}
    \begin{cases}
        0\,, & \text{cases I, IV},\\[4pt]
        \dfrac{B}{B+\mu}\,\chi_\mathrm{e}\alpha\,, & \text{cases II, III},
    \end{cases}
\end{align}
with ${\|\mathbf{Q}\|^{2}=Q_{xx}^{2}+Q_{xy}^{2}}$.

This effective potential acts as a Lyapunov function and allows for a simple graphical interpretation of the dynamics.
The steady states are the stationary points of $V$, and the stability is given by the curvature at the extrema.
The growth rates of the amplitude and rotation modes of the load-aligned fiber states [cf.\ Eq.~\eqref{eq:lsa-decoupled}] are the curvatures of the effective potential at these extrema, ${\lambda_{\parallel}=-\partial^{2}V/\partial Q_{xx}^{2}}$ and ${\lambda_{\perp}=-\partial^{2}V/\partial Q_{xy}^{2}}$.
For general steady states, the growth rates are given by the negative of the eigenvalues of the Hessian matrix.

\begin{figure}
    \centering
    \includegraphics[width=\columnwidth]{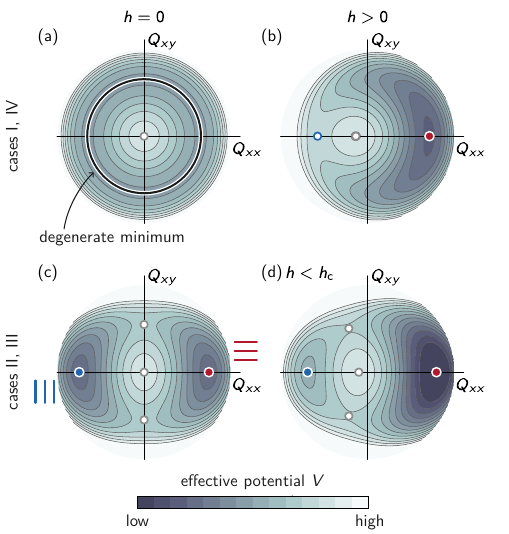}
    \caption{Effective potential $V(Q_{xx}, Q_{xy})$ of the fiber alignment in a homogeneous patch above the alignment threshold [Eq.~\eqref{eq:potential}].
    The effective potential is shown as a contour plot in the ${(Q_{xx}, Q_{xy})}$ plane.
    Darker shading corresponds to lower $V$ (colorbar). Filled circles mark stable states, and open circles denote unstable steady states.
    Red and blue colors mark fibers aligned in the direction $x$ of the external load and perpendicular to it, respectively.
    Top row: cases I and IV, where the elastic-stress feedback does not oppose a rotation of the fibers.
    (a) Case I (no load): the potential has a degenerate ring of minima (black) (cf.\ Sec.~\ref{sec:chie0}).
    (b) Case IV: an applied load tilts the potential, leaving the state with fibers along the load as the only stable steady state.
    (c) Case II: without load, the last term of Eq.~\eqref{eq:potential} breaks the ring of degenerate minima into two orthogonally aligned stable steady states: fiber alignment parallel (red) and perpendicular (blue) to the loading axis $x$.
    (d) case IV: an external load tilts this double-well potential toward the parallel state.
    The perpendicularly aligned steady state becomes shallower and disappears at ${h=h_\mathrm{c}}$.
    }
    \label{fig:potential}
\end{figure}

The potential therefore summarizes our understanding from the previous two sections [Fig.~\ref{fig:potential}].
Its first three terms depend on the fiber direction only through the load. Without load, they form a Mexican hat whose brim is the circle of degenerate aligned states of Sec.~\ref{sec:chie0} [Fig.~\ref{fig:potential}(a)]. The load tilts this brim toward fibers aligned along the loading direction [Fig.~\ref{fig:potential}(b)].
The last term arises from the shear deformation induced by a rotation of the fibers.
It is present only for the mixed boundary conditions II and III, where it penalizes rotations away from the loading axes, corresponding to states with ${Q_{xy}\not=0}$,. This term carves two wells into the brim, separated by barriers at ${Q_{xx}=0}$ [Fig.~\ref{fig:potential}(c)].
Thus, the mechanical feedback from shear strain induces structural bistability between fibers aligned parallel and perpendicular to the load; without it, cases I and IV retain the full brim, and no second state survives. Again, an applied load tilts the landscape. The term ${-h Q_{xx}}$ lowers $V$ at the parallel state, ${Q_{xx} = Q^{+}_{xx} > 0}$, and raises it at the perpendicular state, ${Q_{xx} = Q^{-}_{xx} < 0}$, so that the two wells, degenerate at ${h = 0}$, are no longer of equal depth [Fig.~\ref{fig:potential}(d)].
However, the orthogonally aligned state remains a stable local minimum below the threshold load ${h=h_\mathrm{c}}$, at which the perpendicularly aligned fiber state disappears.
Finally, the landscape shows that the load-aligned and perpendicular fiber states analyzed above are the only steady states that can become stable. Apart from the steady states on the $Q_{xx}$ axis, the effective potential $V$ has saddle points at the barriers in the brim valley.
These saddle points are steady states with tilted fibers, which are always unstable against fiber rotation. 

\section{Domain-wall formation in quasi one-dimensional tissue}
\label{sec:quasi-1D-domain-wall}
To develop an understanding of the effect of the mechanically bistable response in spatially extended tissues, we consider again the quasi one-dimensional tissue patch with periodic boundary conditions in the direction of coordinate $y$, and tissue states that are uniform along $y$ direction. We now analyze the dynamics of the tissue patch without assuming stress, strain, and fiber alignment to be homogeneous.

Given an external pressure $p$ applied in $x$ direction,  the results of Sec.~\ref{sec:chie} show that both parallel and perpendicular fiber alignment $Q_{xx}^* \gtrless 0$ are stable in the homogeneous system.
Thus, we again assume that the fiber alignment is either parallel or perpendicular to the direction of the applied pressure; that is, we set $Q_{xy}=0$, 
Mechanical equilibrium [Eq.~\eqref{eq:force-balance}] requires $\sigma_{xx}$ to be constant throughout the tissue, hence $\sigma_{xx}=-p$.
The periodic boundary conditions require ${\varepsilon_{yy}=0}$, yielding again
\begin{equation}
    \tilde\varepsilon_{xx}(x) = -\frac{1}{2(B+\mu)}\left( p +\alpha Q_{xx}(x)\right),
    \label{eq:hss-quasi-1D-dev-strain}
\end{equation}
as in the homogeneous case, with the only difference that here $Q_{xx}$ and $\tilde\varepsilon_{xx}$ may vary spatially.

Inserting this relation into the fiber dynamics, Eq.~\eqref{eq:fiber-feedback}, one obtains a closed equation for $Q_{xx}$
\begin{align}\label{eq:quasi-1D-AC}
    \partial_t Q_{xx}
    = K \partial_x^2 Q_{xx} + \tilde{a}\, Q_{xx} - 2b\, Q_{xx}^3 - \frac{\chi_\mathrm{e}\mu}{B+\mu}p
    \,.
\end{align}
where
\begin{equation}
    \tilde{a}=\chi_\mathrm{f}\alpha - a -\frac{\chi_\mathrm{e}\alpha \mu}{B+\mu}\,.
\end{equation}
Equation \eqref{eq:quasi-1D-AC} is the Allen--Cahn equation, or bistable-media equation, with a biasing term ${\sim -p}$ \cite{Mikhailov1990,Schlogl1972}. Its local part is a gradient flow, $\partial_tQ_{xx}=-\partial_{Q_{xx}} V$, of the effective potential
\begin{equation}
    V(Q_{xx})=-\frac{\tilde a}{2}Q^2_{xx}+\frac{b}{2}Q^4_{xx}-hQ_{xx},
\end{equation}
whose two minima are the two coexisting aligned states and whose tilt is set by the bias $h$. We will use this representation repeatedly below to read off the local dynamics. 
In the regime of bistability, ${\tilde{a}>0}$
this dynamics forms domain walls of width of order $\ell_\mathrm{dw}= \sqrt{K/\tilde{a}}$ that separate coexisting domains of parallel and perpendicular fibers.
For ${p\lessgtr 0}$, parallel (perpendicular) alignment is preferred, and the domain walls attain the speed \cite{Mikhailov1990}
\begin{equation}
    v = \sqrt{K b}(Q_{xx}^{-}+Q_{xx}^{+}- 2Q_{xx}^{0})
\end{equation}
away from the parallel-aligned domains (in the limit where the domain wall is narrow compared to the system width).
Here, we defined the three roots ${Q_{xx}^{-}-<Q_{xx}^{0}<Q_{xx}^{+}}$ of the bistable ``reaction'' term ${f(Q_{xx})=\tilde{a}\, Q_{xx} - 2b\, Q_{xx}^3 + h}$ with the bias ${h=-\chi_\mathrm{e}\mu\,p/(B+\mu)}$ similar to the discussion in Sec \ref{sec:chie} [see Eq.\eqref{eq:Qpm}].
For positive pressure $p$, the middle root $Q_{xx}^{0}$ moves closer to the positive root $Q_{xx}^{+}$ [Fig.~\ref{fig:flow}] and the velocity becomes negative, and the parallel-aligned domains slowly vanish.
In contrast, the velocity becomes positive for negative pressures ${p<0}$, and the perpendicularly aligned domains shrink.
For a small biasing term, that is, at small pressure, using the approximate values of the roots Eq.\eqref{eq:Qpm} we get ${Q_{xx}^{-}+Q_{xx}^{+}-2Q_{xx}^{0}=3h/\tilde a}$.
Then, the velocity becomes approximately
\begin{equation}\label{eq:quasi-1D-dw-vel}
    v\approx - \frac{3\sqrt{K b}}{\tilde{a}}\frac{\mu}{B+\mu}\chi_\mathrm{e}p\,.
\end{equation}

\section{Bulk Contractility}
\label{sec:bulk-contractility}
So far, we have examined the role of the deviatoric part of the active stress generated by contractile fibers, ${\tilde {\bm\sigma_\mathrm{f}}=\alpha \mathbf{Q}}$.
This term describes a contractile stress exerted along the fiber axis, and an equally strong extensile stress acting in the perpendicular direction. In tissue with a finite bulk modulus, fibers also generate an isotropic or bulk contribution to the active stress. In this section, we discuss the consequences of such an active bulk contraction.

The active stress arises from the sum of the dipolar forces generated by individual contractile units, such as myosin minifilaments, cross-linking actin fibers.

Its general form is given by
\begin{equation}   \bm\sigma_\mathrm{f}=\alpha \mathbf{Q} +\alpha_{\rho}\phi\mathbf{1}\,,
\end{equation}
where ${\alpha_\rho >0}$ is the bulk contractility and $\phi$ the local concentration of contractile units.
The isotropic term acts as a negative active pressure,

${p_{\mathrm{act}}= \alpha_{\rho}\phi}$.
While the isotropic part of the stress does not enter directly in the fiber dynamics, Eq.~\eqref{eq:fiber-dyn}, it does contribute to the force balance, Eq.~\eqref{eq:force-balance}, and thus modifies the elastic-stress feedback.

To analyze the effect of bulk contraction, we need to specify how the concentration $\phi$ evolves in space and time. Motivated by the observation that muscle fibers mature under tension, we expect fiber alignment and concentration to be linked to each other. For simplicity, we therefore assume that the fiber concentration follows the dynamics of the fiber alignment and can be expressed as a function of the degree of order~\cite{braunsActiveSolidsTopological2026, yin2023contractility}
\begin{equation}
    \phi = \phi(||\mathbf{Q}||)\,.
\end{equation}

 We characterize the contractile fibers as two populations: an isotropic background of concentration $\phi_0$ and a fraction organized into aligned fibers. As a minimal choice, we take the concentration of aligned material to be proportional to the magnitude of the nematic order, ${\|\mathbf{Q}\|=\sqrt{\mathrm{tr}[\mathbf{Q}^{2}]/2}}$, setting
\begin{equation}
    \phi=\phi_0+ ||\mathbf{Q}||\,.
      \label{eq:bc-rho}
\end{equation}

A uniform fiber background contributes a constant pressure $p_\mathrm{act}^0={ \alpha_{\rho} \phi_0 }$.
In the quasi-one-dimensional tissue, where $\varepsilon_{yy}=0$, it shifts the external pressure, ${p\xrightarrow{}p + p_\mathrm{act}^0 }$. In both cases, the uniform background contribution can be absorbed into the existing parameters, and we therefore set ${\phi_0=0}$ in the following.  Bulk contractility then acts only where the fibers are ordered.
To determine its role, we re-examine
 
the quasi-one-dimensional tissue patch discussed in Sec.~\ref{sec:quasi-1D-domain-wall}.

We show below that a weak active pressure increases the degree of alignment $||\mathbf{Q}||$ in the direction in which deformations of the tissue are restricted.

\subsection{Domain-wall formation in quasi one-dimensional tissue}
We consider the quasi one-dimensional geometry of Sec.~\ref{sec:quasi-1D-domain-wall}. The tissue is periodic in the $y$ direction and ${\varepsilon_{yy}=0}$.
Moreover, we consider states fulfilling ${Q_{xy}=0}$.
In this case, ${||\mathbf{Q}||= |Q_{xx}|}$, and the active pressure is given by ${p_\mathrm{act}=\alpha_\rho |Q_{xx}|}$.
As before, mechanical equilibrium requires the normal stress to be constant, ${\sigma_{xx}=-p}$. With ${\varepsilon_{yy}=0}$ and the additional isotropic stress $p_\mathrm{act}\mathbf{1}$, we obtain ${\sigma_{xx}=2(B+\mu)\,\tilde\varepsilon_{xx}+\alpha Q_{xx}+p_\mathrm{act}}$, so that the deviatoric strain is
\begin{equation}
  \tilde\varepsilon_{xx}(x)=-\frac{p+p_\mathrm{act}+\alpha Q_{xx}(x)}{2(B+\mu)}\,.
  \label{eq:bc-1D-strain}
\end{equation}
Compared to Eq.~\eqref{eq:hss-quasi-1D-dev-strain}, we see that the active pressure simply adds to  the external pressure $p$. Inserting the deviatoric strain Eq.~\eqref{eq:bc-1D-strain} into the fiber dynamics Eq.~\eqref{eq:fiber-feedback} gives a closed equation for $Q_{xx}$,
\begin{equation}
  \partial_t Q_{xx}=K\partial_x^{2}Q_{xx}+\tilde a\,Q_{xx}-2b\,Q_{xx}^{3}
  - c\,|Q_{xx}|+h\,,
  \label{eq:bc-1D-AC}
\end{equation}

with the bias term ${h=-\chi_\mathrm{e}\mu\,p/(B+\mu)}$ remaining the same as obtained without bulk contraction. 
The active pressure introduces an extra (piecewise) linear coupling ${- c |Q_{xx}|}$ with
\begin{equation} c=\frac{\chi_\mathrm{e}\mu\,\alpha_\rho }{B+\mu}\,.
  \label{eq:bc-c}
\end{equation}
This term is even in $Q_{xx}$ and therefore breaks the symmetry ${Q_{xx}\to-Q_{xx}}$ of the dynamics even in the absence of an external load. It decreases the self-alignment rate of fibers parallel to $x$ and increases the alignment rate of fibers along $y$, with ${\tilde a\to\tilde a\mp c}$ for ${Q_{xx}\gtrless0}$.
Consequently, for ${p=h=0}$, the two homogeneous states have different magnitudes of alignment,
\begin{equation}
  Q^{\pm}_{xx}=\pm\sqrt{\frac{\tilde a\mp c}{2b}}\,.
  \label{eq:bc-1D-amp}
\end{equation}
This asymmetry arises from the constraint $\varepsilon_{yy}=0$. The active pressure is isotropic, but the tissue can deform only along $x$.
Hence, the active pressure compresses the tissue along $x$, generating a contractile deviatoric elastic stress along $y$ which aligns fibers in that direction.
In other words, fibers aligned along $x$ generate an active pressure that suppresses their alignment, while those aligned along  $y$ generate an active pressure that enhances their alignment. The bulk-contraction term therefore favors alignment along the direction in which the tissue cannot deform.

The fiber dynamics still follow bistable-media dynamics, forming a domain wall between the two homogeneous states $Q_{xx}^\pm$.
The broken symmetry ${Q_{xx}\to -Q_{xx}}$ implies that the domain wall obtains a finite velocity at zero bias ${h=0}$. This finite velocity induces domain-wall movement that increases the area of the more strongly ordered domain $Q_{xx}^-$.
A finite, negative external pressure favoring the less ordered domain aligned in $x$-direction is required to suppress domain-wall movement.

\section{Discussion}
Motivated by the emergence of structural order in muscle tissues, we have formulated an active-solid theory of tissues in which contractile fibers self-align on the tissue scale. The tissue is modeled as an elastic medium with embedded contractile fibers that remodel in response to the local deviatoric stresses: the contractile active stress generated by the fibers themselves and the elastic stress of the background tissue.  

The two feedback loops arising from the coupling of fiber alignment to these two contributions to the stress play complementary roles. Feedback from active fiber stress is self-amplifying: fibers contracting along a direction create tension along the same direction, thereby promoting further alignment. As a result,  the tissue orders spontaneously once the contractility exceeds a threshold. Feedback from the elastic stress couples the fiber alignment to geometry and boundary conditions through force balance that controls the mechanical response of the tissue. Together, these two feedback mechanisms generate both structural and mechanical patterning of the tissue.

Our first central finding is that mechanical boundary conditions, via elastic-stress feedback, select two orthogonal fiber directions and determine their stability. With fiber-stress feedback alone, all alignment directions are equivalent [Fig.~\ref{fig:potential}(a)]. An applied load breaks this rotational symmetry and selects stationary alignment in the directions parallel and perpendicular to the load. Alignment along the principal extensile load axis is always stable.
Additionally, alignment perpendicular to the load axis is stable as well.
Only then does the shear strain induced by a rotation of the fibers oppose this rotation and stabilize perpendicular alignment, by building a 'hill' of effective potential between the parallel and perpendicular states [Fig.~\ref{fig:potential}(c) and Eq.~\eqref{eq:potential}]. The result is structural and mechanical bistability of the tissue, resulting in a hysteretic stress--strain response.

Taken together, our results elucidate a minimal mechanism for patterning structural order into orthogonally aligned regions. Orthogonal fiber alignment is ubiquitous in biological systems, and it is an intriguing future task to test whether the proposed mechanism might underlie the orthogonal alignment of fiber layers in different experimental systems.
Extending our theory to elastically coupled tissue layers in three-dimensional geometries would test whether the same mechanical-feedback mechanism might explain perpendicular muscle alignment in such tangential layers. We address this question for the two-layered tissue of \textit{Hydra} in a forthcoming publication \cite{Weyer.Braunsinpreparation}. 
Moreover, a remarkable example of structural patterning of muscle tissue is the adult human heart~\cite{streeter1969fiber}. The intricate helical arrangement of muscle fibers around the left ventricle demands a better understanding of how geometry guides structural patterning.
The effect of stress feedback in fully three-dimensional active elastic tissue is thus a promising direction for future research, all the more given the diversity of geometries in which muscle tissues develop across organisms. A three-dimensional extension of our model to biologically motivated geometries may reveal a multitude of fiber patterns regulated by mechanical feedback and geometry. We study this in a forthcoming publication. 

Moreover, we have considered the quasi-static limit of the mechanics. This limit is appropriate for muscle tissues that sustain tonic load, such as many smooth muscles, and we expect it to qualitatively describe the average response of a tissue under cyclic contraction and relaxation, like the cardiac muscle. Future work will couple oscillatory signaling and contraction to mechanics to understand how cyclic contraction affects tissue patterning and how patterning may feed back onto the synchronization of oscillatory chemical signaling.

Finally, we have described the background tissue with a fixed reference configuration by treating the background as linearly elastic. This description applies to muscle tissues that order structurally without changing their shape significantly. Many biological tissues, however, relax stress over long timescales through cell rearrangement, cell division, and extracellular matrix turnover. Therefore, incorporating plastic or viscoelastic deformations would allow for the reference state to be remodeled, leading to slowly evolving patterns and shape morphogenesis \cite{yuFeedbackcontrolledEpithelialMechanics2026}.
Indeed, anisotropic force generation due to aligned actomyosin underlies key morphogenetic processes like Drosophila germband extension \cite{streichanGlobalMorphogeneticFlow2018} or chick gastrulation \cite{Caldarelli.etal2024, serraMechanochemicalModelRecapitulates2023}
A similar mechanical feedback has been argued to contribute to epithelial gap closure \cite{perez-verdugoMechanosensitiveFeedbackOrganizes2026}. 
More generally, our work shows that stress feedback can generate structural patterning by coupling with the geometry of the tissue. In our model, no chemical or morphogenetic pattern is needed to set the fiber direction; the boundary conditions select them, and global force balance positions the domains. 

\section*{Acknowledgments}
We thank Fridtjof Brauns, Markus Merk, and Sebastian Streichan for insightful discussions. We also thank Toshi Parmar, Dimitrios Krommydas, and other members of the Marchetti and Streichan groups for helpful discussions. 
This work was supported by NSF award DMR-2528734.
H.W. was supported by grant GBMF7392 from the Gordon and Betty Moore Foundation and NSF PHY-2309135 to the Kavli Institute for Theoretical Physics (KITP). This research has benefited from the Mitchel Postdoctoral Scholar Career Development Fund. We acknowledge the use of Claude (Opus 5.5) for coding, preliminary calculations, and editing.

\end{document}